\documentstyle[12pt,epsfig]{article}
\newlength{\pubnumber} 

\catcode`\@=11
\@addtoreset{equation}{section}
  
\def\section{\@startsection{section}{1}{\z@}{3.5ex plus 1ex minus .2ex}
 {2.3ex plus .2ex}{\large\bf}}
\def\subsection{\@startsection{subsection}{2}{\z@}{2.3ex plus .2ex}
 {2.3ex plus .2ex}{\bf}}

\newcommand{\gsim}{\lower.7ex\hbox{$\;\stackrel{\textstyle>}{\sim}\;$}}
\newcommand{\lsim}{\lower.7ex\hbox{$\;\stackrel{\textstyle<}{\sim}\;$}}

\begin{document}

\begin{titlepage}
\samepage{
\setcounter{page}{1}
\rightline{RAL-TR-1999-052; TPI--MINN--99/35; UMN--TH--1809--99}
\rightline{July 1999, {\tt hep-ph/9907331}}
\vfill
\begin{center}
 {\Large \bf  $D$--term Spectroscopy in \\Realistic Heterotic--String Models\\}
\vfill
 {\large                    Athanasios Dedes$^1$\footnote{
   E-mail address: A.Dedes@rl.ac.uk} 
                $\,$and$\,$ Alon E. Faraggi$^{2,3}$\footnote{
   E-mail address: faraggi@mnhepo.hep.umn.edu}\\}
\vspace{.09in}
 {\it  $^{1}$ Rutherford Appleton Laboratory,
  Chilton, Didcot, Oxon, OX11 0QX, UK\\}
\vspace{.025in}
 {\it  $^{2}$Theoretical Physics Institute,
  University of Minnesota,
  Minneapolis, MN 55455\\}
\vspace{.025in}
{\it $^{3}$  Theory Division, CERN, 
                1211 Geneva, Switzerland \\}
\end{center}
\vfill
\begin{abstract}
  {\rm
The emergence of free fermionic string models
with solely the MSSM charged spectrum below the string scale
provides further evidence to the assertion that the true string vacuum
is connected to the $Z_2\times Z_2$ orbifold in
the vicinity of the free fermionic point in the Narain
moduli space. An important property of the $Z_2\times Z_2$
orbifold is the cyclic permutation symmetry between 
the three twisted sectors. If preserved in the three generations
models the cyclic permutation symmetry results 
in a family universal anomalous $U(1)_A$, which is instrumental
in explaining squark degeneracy, provided that the dominant
component of supersymmetry breaking arises from the $U(1)_A$
$D$--term. Interestingly, 
the contribution of the family--universal $D_A$-term to the
squark masses may be intra-family non-universal,
and may differ from the usual (universal) boundary conditions assumed
in the MSSM. We contemplate how $D_A$--term spectroscopy 
may be instrumental in studying superstring models
irrespective of our ignorance of the details of supersymmetry breaking.
We examine the possible effect of the intra--family non--universality 
on the resulting SUSY spectrum and the values of the
strong coupling, effective weak mixing angle and W-gauge boson 
mass, up to a two loop accuracy, in the two models
(universal and non-universal). We find that non universality
relaxes the constraint of color and charge breaking minima which
appears in the universal case. In addition, it predicts a 3\% smaller
value of $\alpha_s$ due to different threshold masses obtained 
in the latter scenario. Finally, we present the experimentally allowed 
predictions of the two models in an $M_0$ and $M_{1/2}$ parameter space.
}
\end{abstract}
\vfill
\smallskip}
\end{titlepage}

\setcounter{footnote}{0}

\def\beq{\begin{equation}}
\def\eeq{\end{equation}}
\def\beqn{\begin{eqnarray}}
\def\eeqn{\end{eqnarray}}
\def\AEF{A.E. Faraggi}
\def\NPB#1#2#3{{\it Nucl.\ Phys.}\/ {\bf B#1} (19#2) #3}
\def\PLB#1#2#3{{\it Phys.\ Lett.}\/ {\bf B#1} (19#2) #3}
\def\PRD#1#2#3{{\it Phys.\ Rev.}\/ {\bf D#1} (19#2) #3}

\def\PRL#1#2#3{{\it Phys.\ Rev.\ Lett.}\/ {\bf #1} (19#2) #3}
\def\PRT#1#2#3{{\it Phys.\ Rep.}\/ {\bf#1} (19#2) #3}
\def\MODA#1#2#3{{\it Mod.\ Phys.\ Lett.}\/ {\bf A#1} (19#2) #3}
\def\IJMP#1#2#3{{\it Int.\ J.\ Mod.\ Phys.}\/ {\bf A#1} (19#2) #3}
\def\nuvc#1#2#3{{\it Nuovo Cimento}\/ {\bf #1A} (#2) #3}
\def\etal{{\it et al,\/}\ }
\hyphenation{su-per-sym-met-ric non-su-per-sym-met-ric}
\hyphenation{space-time-super-sym-met-ric}
\hyphenation{mod-u-lar mod-u-lar--in-var-i-ant}


\section{Introduction}

Superstring phenomenology aims at achieving two goals.
The first task is to reproduce the phenomenological
data provided by the Standard Particle Model. The
subsequent goal is to extract possible experimental
signatures which may provide further evidence for 
the validity of specific string models, in particular,
and for string theory, in general. 

The most realistic superstring models constructed to date
are those in the free fermionic formulation \cite{revamp,fny,alr,
eu,top,cfn}. 
Not only do these models naturally give rise to three
generations with the $SO(10)$ embedding of the
Standard Model spectrum, but it was recently also shown,
that free fermionic models
can also produce models with solely the MSSM charged spectrum
below the string scale \cite{cfn}. Thus, for the first time
we have an example of a Minimal
Superstring Standard Model{\bf !}
The success of the free fermionic models suggests that some of 
their underlying structure will persist in the
true string vacuum. The key properties, which may be 
the origin of the phenomenological success of the 
free fermionic models, are: 1) the fact that the
free fermionic formulation is formulated at an enhanced
symmetry point in the Narain moduli space; and 2)
their relation with $Z_2\times Z_2$ orbifold compactification,
which underlies the free fermionic models. 
The phenomenological success of the free fermionic models
provides evidence to the assertion that the true
string vacuum is connected to the $Z_2\times Z_2$
orbifold in the vicinity of the free fermionic point
in the Narain moduli space. 

Subsequent to establishing the phenomenological viability
of heterotic--string
free fermionic models we may seek possible experimental signatures
which will provide further evidence to the validity of specific models,
in particular, and to string theory, in general.
One such possible signature which has been discussed in the past
is the appearance of exotic states with fractional $U(1)_Y$
or $U(1)_{Z^\prime}$ charge \cite{ccf}. Such states appear because
of the breaking of the non--Abelian gauge symmetries
by ``Wilson--lines'' in string theory. While on
the one hand the existence of such light states
imposes severe constraints on otherwise valid
string models \cite{lykken}, provided that the exotic states
are either confined or sufficiently heavy,
they can give rise to exotic signatures. For example,
they can produce heavy dark matter candidate, possibly
with observable consequences \cite{ben}. 

In this paper we discuss another possible signature
of realistic string models. Realistic string models
possess $N=1$ space--time supersymmetry. Different
mechanism for breaking supersymmetry have been proposed. 
These include the ideas of : (i) gaugino condensation in
the hidden sector \cite{gauginocond}; (ii) 
dilaton dominated SUSY breaking  \cite{DDSB};
(iii) gauge mediated SUSY breaking \cite{GMSB}; and (iv) SUSY breaking
induced by an anomalous $U(1)$ $D$--term together with effective
mass term of certain relevant fields  \cite{fh1,fp2}. 
A vital issue in SUSY phenomenology is the origin of the extreme
degeneracy in the masses of the squarks in at least the first
two families as inferred from the minuscule strengths of the
$K^0-{\bar K}^0$ transition. The problem becomes especially
acute when considering theories which consistently
unify gravity with the gauge interactions. For example,
in string theory the soft SUSY breaking terms are in general
expected to be not family universal \cite{il}. String models that
may explain the required mass degeneracy are therefore
especially interesting. Recently it was shown that free 
fermionic models possess the desired structure
to explain the required squark degeneracy \cite{fp2,cf1}. The
important feature is the relation of the free
fermionic models to $Z_2\times Z_2$ orbifold
compactification, which possesses a cyclic permutation
symmetry between the three twisted orbifold sectors.
In some of the three generation models this cyclic
permutation symmetry is preserved \cite{fp2,cf1}. The permutation
symmetry is reflected in the charges of the three 
generations under the horizontal $U(1)$ symmetries,
resulting in the anomalous $U(1)$ being family universal.
In the case that the family--universal anomalous $U(1)$ provides 
the dominant source of SUSY breaking, the
squark masses are family universal. 
The interesting aspect in regard to the
anomalous $U(1)$ charges is that, although
they are family universal, they may have intra--family
non--universal charges. In this case, although
the contribution of the anomalous $U(1)$ $D$--term
to the squark masses is family universal, it is 
intra--family non--universal, and differs
from the usual boundary conditions assumed
in the MSSM. Consequently, the resulting
sfermion spectrum will have a distinctive signature
which differs from that of the MSSM. Furthermore,
suppose that there are several sources which
contribute to the sfermion masses. Some of these
sources may be family and intra--family universal,
like the one arising from the dilaton. On the other
hand, the anomalous $U(1)$ $D$--term may contribute
a family universal, but intra--family non--universal
component to the sfermion masses. It is this
component which one would like to extract from
the supersymmetric spectrum in future experiments.

In this paper we examine these ideas in the framework
of the free fermionic superstring models. For concreteness
we focus on two of the standard--like models. One
which produces family and intra--family universal squark masses,
and the second which produces family universal
but intra--family non--universal sfermion masses. 
We contemplate how $D_A$--term spectroscopy 
may be instrumental in studying superstring models
irrespective of our ignorance of the details of supersymmetry breaking.
We examine the possible effect of the intra--family non--universality 
on the resulting SUSY spectrum and the values of the
strong coupling, effective weak mixing angle and W-gauge boson 
mass, up to a two loop accuracy, in the two models
(universal and non-universal). We find that non universality
relaxes the constraint of color and charge breaking minima which
appears in the universal case. In addition, it predicts a 3\% smaller
value of $\alpha_s$ due to different threshold masses obtained 
in the latter scenario. Finally, we present the experimentally allowed 
predictions of the two models in an $M_0$ and $M_{1/2}$ parameter space.

\section{Anomalous $U(1)$ SUSY breaking in free fermionic models}

Let us recall that a model in the
free fermionic formulation is defined by a set of boundary
condition basis vectors, and the associated one--loop
GSO projection coefficients \cite{fff}. 
The massless spectrum is obtained by applying the generalised
GSO projections. A physical state defines a vertex operator
which encodes all the quantum numbers with respect to the 
global and gauge symmetries. Superpotential terms are then
obtained by calculating the correlators between the
vertex operators \cite{cvetic,kln}.

The realistic free fermionic models
are constructed in two stages. The first stage consists
of the NAHE set, $\{{\bf1}, S, b_1,b_2,b_3\}$. This 
set of boundary condition basis vectors has
been discussed extensively in the literature \cite{nahe}.
The properties of the NAHE set are important
to understand the emergence of a family universal anomalous
$U(1)$ \cite{cf1}.
The gauge group after imposing the GSO projections of the NAHE
set basis vectors is $SO(10)\times SO(6)^3\times E_8$.
The three sectors $b_1$, $b_2$ and $b_3$ produce
48 multiplets in the chiral 16 representation of $SO(10)$.
The states from each sector transform under the flavor,
right--moving $SO(6)_j$ gauge symmetries, and under the
left--moving global symmetries.
The cyclic permutation symmetry between the basis vectors
$b_1$, $b_2$ and $b_3$ is the root cause
for the emergence of flavor universal anomalous $U(1)$
in some free fermionic models. 
This is further exemplified by adding to the
NAHE set the boundary condition basis vector $X$ \cite{foc}.
With a suitable choice of the generalised GSO projection
coefficients, the $SO(10)$ gauge group is enhanced to $E_6$.
The $SO(6)^3$ symmetries are broken to $SO(4)^3\times U(1)^3$.
One combination of the $U(1)$ symmetries is embedded in $E_6$,
\beq
U(1)_{E_6}={1\over\sqrt{3}}(U_1+U_2+U_3).
\label{u1e6}
\eeq
This $U(1)$ symmetry is flavor
independent, whereas the two orthogonal combinations
\beqn
U(1)_{12} &=& {1\over\sqrt{2}}(U_1-U_2)~~~~;\label{u12}\\
U(1)_\psi &=& {1\over\sqrt{6}}(U_1+U_2-2U_3)\label{upsi}
\eeqn
are flavor dependent.
The final gauge group in this case is therefore
$E_6\times U(1)^2\times SO(4)^3\times E_8$. 

In the realistic free fermionic models the $E_6$ symmetry is replaced by 
$SO(10)\times U(1)$. This can be seen to arise either by substituting
the vector $X$, with a boundary condition basis vector $2\gamma$ \cite{foc};
or by the choice of the GSO phase $c(X,\xi)=\pm1$,
where $\xi={\bf1}+b_1+b_2+b_3$.
In both cases the right--moving gauge group
is $SO(10)\times U(1)_A\times U(1)^2\times SO(4)^3\times
SO(16)$.
The $E_6\times E_8$ gauge group
in both cases is replaced by $SO(10)\times U(1)_A\times SO(16)$
where $U(1)_A$ is the anomalous $U(1)$ combination. We therefore
see how in this case the anomalous $U(1)$ is just the combination
which is embedded in $E_6$ and its flavor universality is fact 
arises for this reason. 

The NAHE set and the related
$E_6\times E_8$ and $SO(10)\times U(1)_A\times SO(16)$ models
are the first stage in the construction of the 
three generation free fermionic models. The next step
is the construction of several additional boundary condition basis
vectors. These additional boundary condition basis vectors
reduce the number of generations to three generations,
one from each of the sectors $b_1$, $b_2$ and $b_3$.
The additional boundary condition basis vectors break
the $SO(10)$ gauge group to one of its subgroups and
similarly for the hidden $SO(16)$ gauge group. At the same time
the flavor $SO(4)^3$ symmetries are broken to factors of $U(1)$'s.
The number of these $U(1)$'s depends on the specific
assignment of boundary conditions for the set of internal
world--sheet fermions and can vary from 0 to 6.
At the level of the $SO(10)\times U(1)_A\times SO(4)^3$ model
there exist a permutation symmetry between the sectors
$b_1$, $b_2$ and $b_3$ with respect to their charges
under the $SO(4)^3$ symmetries. When the $SO(4)^3$ symmetries
are broken to factors of $U(1)$'s this permutation symmetry
will in general be broken. It is remarkable, however,
that in some of the three generation models the
permutation symmetry between the sectors $b_1$, 
$b_2$ and $b_3$ with respect to their charges 
under the horizontal $U(1)$ symmetries is retained.
In those cases the anomalous $U(1)$ combination is
family universal. In the model of ref. \cite{top}
the anomalous $U(1)$ is just the combination in eq. (\ref{u1e6}),
whereas the two orthogonal combinations are those
in eqs. (\ref{u12}) and (\ref{upsi}).
In the model of ref. \cite{top} the charges of the 
anomalous $U(1)$ charges of the three generations
are both family universal and intra--family universal. 

The standard--like model of ref. \cite{eu} and the flipped
$SU(5)$ model of ref. \cite{price} exhibit a similar structure
of the anomalous $U(1)$ and anomaly free combinations.
In these two models the $U(1)$ symmetries, generated by the world--sheet
complex fermions $\{{\bar\eta}^1,{\bar\eta}^2,{\bar\eta}^3\}$ and
$\{{\bar y}^3{\bar y}^6,{\bar y}^1{\bar\omega}^5,{\bar\omega}^2
{\bar\omega}^4\}$ (or $\{{\bar y}^4{\bar y}^5,{\bar y}^1{\bar\omega}^6,
{\bar\omega}^2{\bar\omega}^3\}$ are anomalous, with:
${\rm Tr} U_1=
{\rm Tr} U_2={\rm Tr} U_3=24,{\rm Tr} U_4= {\rm Tr} U_5=
{\rm Tr} U_6=-12$. 
The anomalous $U(1)$ combination in both models is therefore given by
\beq
U_A={1\over{\sqrt{15}}}(2 (U_1+U_2+U_3) - (U_4+U_5+U_6))~;~ {\rm Tr} Q_A=
{1\over{\sqrt{15}}}180~.
\label{u1a}
\eeq
One choice for the five anomaly--free combinations is
given by
\beqn
{U}_{12}&=& {1\over\sqrt{2}}(U_1-U_2){\hskip .5cm},{\hskip .5cm}
{U}_{\psi}={1\over\sqrt{6}}(U_1+U_2-2U_3),\label{u12upsi}\\
{U}_{45}&=&{1\over\sqrt{2}}(U_4-U_5){\hskip .5cm},{\hskip .5cm}
{U}_\zeta ={1\over\sqrt{6}}(U_4+U_5-2U_6),\label{u45uzeta}\\
{U}_\chi &=& {1\over{\sqrt{15}}}(U_1+U_2+U_3+2U_4+2U_5+2U_6).
\label{uchi}
\eeqn

The anomalous $U(1)$ in the model of ref. \cite{eu} is family universal,
but is intra--family non--universal. This arises because
of the charges of the three generations under the three 
horizontal symmetries $U(1)_{4,5,6}$. Although the permutation symmetry
between the sectors $b_1$, $b_2$ and $b_3$ with respect
to charges under these three $U(1)'s$ is maintained,
the charges differ between members of each family.

%

Supersymmetry breaking in the presence of a
family--universal anomalous $U(1)$ symmetry 
in the realistic free fermionic models was analysed
in detail in ref. \cite{fp2}.
Supersymmetry breaking will occur,
at hierarchically small scale if there is a mass term,
$m\Phi{\bar\Phi}$,
for some Standard Model singlet, which is charged under the
anomalous $U(1)$. The effective potential then takes the form
\begin{equation}
V={g^2\over2}\sum_\alpha D_\alpha^2+m^2(\vert\Phi\vert^2
+\vert{\bar\Phi}\vert^2)
\end{equation}
where $D_\alpha$ are the various $U(1)$ $D$--terms,
and we assumed a common coupling $g$ at the unification scale,
to simplify the analysis. Extremizing the potential
it is found that SUSY is broken. Furthermore, for a specific 
solution of the $F$ and $D$ flatness constraints
it is found that the mass term $m$ is hierarchically suppressed
and that in the minimum the $D$--terms of the family universal
$U(1)$'s are nonzero, whereas those of the family dependent
$U(1)$ vanish. This solution therefore provides an example
how the squark mass degeneracy may arise, provided that the
dominant component that breaks supersymmetry is the anomalous
$U(1)$ $D$--term. 
Furthermore, the mass term $m$, which breaks supersymmetry,
can be hierarchically small relative to the Planck scale.
This is because such a term must arise from
nonrenormalizable terms that contain hidden sector
condensates. The condensation scale in the hidden sector
is determined by its gauge and matter content.
For example, in the model of ref \cite{eu} we found a cubic
level flat $F-D$ solution, with the mass term $m$ induced
at order $N=8$, by matter condensates of the hidden $SU(5)$
gauge group \cite{fp2}.
A numerical estimate of the mass term $m$ yielded $m\sim(1/2-50){\rm TeV}$. 
The analysis of flat directions and minimisation of
of the potential in the presence of the mass term 
was performed in ref. \cite{fp2} for the
string models of ref. \cite{top} and \cite{eu}.
The important aspect is the
distinction between the two models with respect
to the charges of the chiral generations under the
anomalous $U(1)$ symmetry. In the model of ref. 
\cite{top} the anomalous $U(1)$ combination is
given in eq. (\ref{u1e6}) and is both family universal 
and intra--family universal. On the other hand
in the model of ref. \cite{eu} the anomalous 
$U(1)$ combination is given in eq. (\ref{u1a})
and is family universal but not intra--family universal.
The contribution of the anomalous $U(1)$ $D$--term
to the squark masses is given by 
\beq
[m^2_{{\tilde q}_i}]_{D_A}=g^2Q_A^i\langle D_A\rangle,
\label{squaku1amasses}
\eeq
and likewise for the sleptons. Here $Q_A^i$ are the charges 
of the sfermions under the anomalous $U(1)$ and $\langle D_A\rangle$
is the vev of the $D$--term of the anomalous $U(1)$ in the 
minimum of the potential. Thus, assuming that the anomalous
$U(1)$ provides the dominant source of supersymmetry breaking,
the two models will yield different boundary conditions
for the soft SUSY breaking terms at the unification scale. 
With this assumption, whereas the model of ref. \cite{top}
produces the usual family and intra--family universal
boundary conditions for the soft SUSY sfermion masses 
\beq
\left[m^2({\tilde Q}_L):m^2({\tilde u}_R):m^2({\tilde d}_R):
m^2({\tilde L}):m^2({\tilde e}_R)\right]_{D_A}=~1:1:1:1:1~~~
\label{ratiosol2}
\eeq
the model of ref. \cite{eu} produces the boundary
conditions which are family universal but intra--family
non--universal boundary conditions,
\beq
\left[m^2({\tilde Q}_L):m^2({\tilde u}_R):m^2({\tilde d}_R):
m^2({\tilde L}):m^2({\tilde e}_R)\right]_{D_A}=~3:1:3:1:1~~~
\label{ratiosol1}
\eeq

\section{Superstring $D$--term spectroscopy}

The boundary conditions in Eq. (\ref{ratiosol2}) and (\ref{ratiosol1})
represent the contribution of the anomalous $U(1)$ $D$--term
to the sfermion masses. As argued in ref. \cite{fp2}
it is likely that this contribution will be accompanied
by another coming, for example, from the dilaton VEV.
Thus, the soft SUSY breaking boundary conditions may
include a piece which is family and intra--family
universal as well as the anomalous $U(1)$ $D$--term
contribution which is family universal but
may be intra--family non--universal.
The important point is that in the superstring models the charges under
the anomaly free and anomalous $U(1)$ symmetries
are given. Thus, in the event that future experiments
observe the supersymmetric partners, specific patterns
of the observed SUSY spectrum will be correlated
with specific patterns of charges in the superstring models.
Naturally, a full correlation will require a more
complete solution to the problem of supersymmetry
breaking in string theory. Nevertheless, it is
obvious that at first attempt what will be 
required is a crude analysis of the type that we
discuss here. Furthermore, the phenomenological data
to be provided by the future SUSY spectrum will
be instrumental in constraining the viable
superstring models. 
Suppose then that at the unification scale 
the soft SUSY breaking parameters are give by a piece
which is family and intra--family universal
as well as one which depends on the anomalous $U(1)$
$D$--term. 
It is precisely the piece 
which depends on the anomalous $U(1)$ charge
which we will want to extract in future experiments.
In our analysis below we will assume heuristically
that the soft SUSY parameters are given at the MSSM
unification scale, and a more refined analysis will
have to address the issue of bridging the MSSM and
string unification scale, either by the inclusion
of additional matter states \cite{price,matter}
or by Witten's M--theory solution \cite{witten}.
With this assumptions the scalar masses at the
low scale are parametrised by the usual $m_{1/2}$, $m_0$, 
and $A$, soft SUSY breaking parameters, and 
the two Higgs mixing parameters, as well as the
anomalous $U(1)$ $D$--term contribution. 
Using the renormalization group equations (RGE's), the soft supersymmetry
breaking masses at low energies are
calculated from the parameters at the unification scale.

Although Yukawas contribute to sparticle masses in the RGE's 
and their effect is
numerically calculable in terms of the KM angles and the
top quark mass, it
is convenient to eliminate their influence from this program \cite{kelley}.
First, we may safely neglect all but the top and bottom quark
Yukawas.  Second, in a charge-$2/3$ quark
mass eigenstate superfield basis, the
top quark Yukawa will contribute only to third--generation sparticle masses.
The bottom Yukawa will give non--diagonal contributions involving
the first and second generations which will lead to the requirement of
an explicit
diagonalization of the $6\times 6$ up and down 
squark mass matrices.  However, these off--diagonal 
contributions are suppressed by KM angles mixing the third
generation to the first and second, and except for very large
values of $tan\beta$ (which imply a large bottom Yukawa) may be
neglected.  In the case of equal top and
bottom Yukawas at the unification scale, 
the effects of Yukawas can be 
included in a full numerical analysis.
But for the purposes of the discussion here
we restrict our analysis to first and second generation
sparticles and neglect the generally small effects of Yukawas on these masses.
This has the additional advantage of removing the
soft supersymmetry breaking trilinear coupling $A$ and the superpotential
Higgs mixing parameters $\mu$ and $B$ from the analysis.  
Because of experimental difficulties in detecting neutral particles,
and the possibility of confusion between the many other neutral
particles in supersymmetric theories, we also eliminate the
sneutrinos from our phenomenological discussion.

Under these assumptions the light--generation 
sparticle masses may then be analytically calculated from the
one--loop RGE's in terms of the three unknowns $m_{1\over2}$, $m_0$,
$cos2\beta$, and the VEV of the anomalous $U(1)_A$ $D$--term,
$\langle D_A\rangle$, which is of the order of the electroweak
scale \cite{fp2}:
\beq
m^2_{\tilde p}={\tilde m}^2_{0}+
c_{\tilde p}m^2_{1\over2}+
d_{\tilde p}cos2\beta M^2_W +Q_A^{\tilde p}\langle D_A\rangle
\label{spm}
\eeq
where ${\tilde m}^2_{0}$ contain all the family
universal contributions, like those arising from the
dilaton VEV and $Q_A^{\tilde p}$ is the charge of a sparticle
under $U(1)_A$.
The coefficients $c_{\tilde p}$ for the different sparticles
result from the running of the gaugino masses, and 
$d_{\tilde p}=2(T_{3_L}^{\tilde p}-{3\over5}Y^{\tilde p}\tan^2\theta_W)$
results from the electroweak Higgs VEVs. The last piece
entails the anomalous $U(1)$ $D$--term contribution, and we
absorbed all universal factors into $\langle D_A\rangle$. It 
is this last piece that we would want to extract 
from a future supersymmetric spectrum, as it depends
on the specific $U(1)_A$ charges in a given string model.
For example, given the ratio of $U(1)_A$ charges in eq. 
(\ref{ratiosol1}) we are interested in extracting the
relative weight between the different family members.
Then we can absorb all family universal dependence
into $\langle D_A\rangle$. 
The resulting equations (\ref{spm}) will then depend on
the anomalous $U(1)_A$ charges of the various sparticles.
The equations can then be solved for $\langle D_A\rangle$
and through their dependence on the charges
different models will produce distinctive
dependence on the measured sparticle masses. 
For example, with the charges given in Eq. (\ref{ratiosol1})
we have
\beqn
\cos2\beta &=& {{(m^2_{\tilde u_l}-m^2_{\tilde d_l})}\over{2M_W^2}}=
{{\Delta Q}\over{2M_W^2}}\nonumber\\
m^2_{1/2} &=& {{(m^2_{\tilde d_l}-m^2_{\tilde d_r})-
(d_{\tilde d_l}-d_{\tilde d_r}){{\Delta Q}\over{2}}}\over
{(c_{\tilde d_l}-c_{\tilde d_r})}}\nonumber\\
\langle D_A\rangle &=& {1\over2}\left((m^2_{\tilde d_r} - m^2_{\tilde u_r})-
(c_{\tilde d_r} -c_{\tilde u_r})
m^2_{1/2}- (d_{\tilde d_r} - d_{\tilde u_r}){{\Delta Q}\over2}\right)
\label{cos2bm12}
\eeqn
and similarly for $m_0^2$. Thus, four measured sparticle
masses can be used to test the specific hypothesis on the 
source of the soft SUSY breaking terms, Eq. (\ref{ratiosol1}).
More generally, the measured sparticle masses will
be used to investigate their correlation with the
charges in specific string models.
Such hypothesis will then be further tested by the additional
sfermion masses. Just as the Standard Model charges 
provide strong support for an underlying $SO(10)$ structure,
a successful correlation will provide further evidence for
such successful string models. We shall postpone a more 
detailed analysis of the sfermion spectroscopy in
these models until the supersymmetric spectrum is
actually observed. In the next section we
will examine the possible effect of the
non--universal stringy boundary conditions 
on Z--scale observables. 

\section{Spectroscopy and Z--Observables}

In this section we examine the possible effect of
the string intra--family nonuniversal boundary
conditions on Z--scale observable. For concreteness we assume that the
boundary conditions are given at the MSSM unification scale
and extrapolate to low energy assuming the MSSM spectrum. 
More detailed study, including the effect of additional matter,
is delegated to future work. 
In the following, we make a numerical analysis of the 
two cases we mentioned so far :

\begin{center}
{\bf Universal :}
\end{center}
\beqn
 m^2(\tilde{Q}_L): m^2(\tilde{u}_R):
 m^2(\tilde{d}_R):  m^2(\tilde{L}):  m^2(\tilde{e}_R) =M_0^2 :
M_0^2 :M_0^2 :  M_0^2 :M_0^2  \;, \nonumber \\
\label{un}
\eeqn
and

\newpage

\begin{center}
{\bf Non Universal :}
\end{center}
\begin{eqnarray}
m^2(\tilde{Q}_L) : m^2(\tilde{u}_R):
 m^2(\tilde{d}_R) :  m^2(\tilde{L}) :  m^2(\tilde{e}_R) =3\times M_0^2 :
M_0^2 :3\times M_0^2 :  M_0^2 :M_0^2  \;. \nonumber \\
\label{nonun}
\end{eqnarray}

We use two loop Renormalization group equations for the evolution of  every 
coupling and mass appeared in the model. In fact, we start
by defining the gauge couplings using the most precise 
experimental quantities : the Fermi coupling constant 
$G_F=1.16639 \times 10^{-5}~ {\rm GeV}^{-2}$, the electromagnetic coupling
$\alpha_{EM}(1~ {\rm GeV})=1/137.036$ and 
 the Z-boson mass, $M_Z=91.187 ~{\rm GeV}$. These three quantities can
be used to define the running value of the weak mixing angle (following 
the analysis of Refs.\cite{Dedes:1998hg,Dedes:1997wc}) and thus 
the running gauge couplings $g_1=\sqrt{\frac{5}{3}}\frac{e}{\cos\theta_W}$ and 
$g_2=\frac{e}{\sin\theta_W}$. We evolve them up to the scale (GUT scale)
 where the couplings meet and we set the value of the
$g_3$ equal to $g_{GUT}=g_1=g_2$. At this scale we impose the 
universal (non-universal) boundary conditions of eqs. (\ref{un},\ref{nonun}).
We run all the parameters down to the EW scale by assuming Radiative
Electroweak symmetry breaking, where  the full 1-loop
contributions to the minimisation conditions of the effective potential have
been included. Note also that in the above scheme both finite 
and logarithmic threshold effects are taken properly into account
~\cite{Faraggi:1994qb,Bagger:1995bw,Dedes:1997wc}.
We treat the thresholds for every mass which appeared in the model by
using the so called 'theta'-function approximation
\cite{Lahanas:1995dj,Dedes:1996sb}. That is when a running
mass $m(Q)$ passes through its physical mass which is defined 
as $m(Q)=Q$, then  this mass gets decoupled from the rest of the 
RGE's.  Convergence with the above boundary conditions  is reached
after few iterations and the outputs contain :
the strong QCD coupling, $\alpha_s(M_Z)$, the (leptonic) 
effective weak  mixing angle, $s_{eff}^{lep}(M_Z)$ ({\it i.e.,} see
Ref.\cite{Dedes:1998hg,Dedes:1997wc} for more  details) the W-pole mass and
the sparticle spectrum. 

In Table I we review the current experimental bounds on the 
SUSY and Higgs particles, we have made use in this analysis.
We also display the (theoretical) assumptions which 
have been used in  the derivation of these bounds. The
references of the most recent  relevant articles are also
displayed.

 In Fig.\ref{fig1}
we display the excluded regions in both cases of universal (M-SUGRA)
and non-universal boundary conditions and for two rather extreme
values of the $\tan\beta$.
Clearly, non-universality relaxes
some of the experimental bounds.  Thus, in the case of Universal
boundary conditions the parameter space with $M_0 \lsim 500$ GeV and 
$M_{1/2} \lsim 190$ GeV is ruled out by the Higgs searches\footnote{We
have used one loop corrections for the evaluation of the light Higgs boson
 mass.} 
while in the case of the non-universal
boundary conditions the corresponding excluded region is $M_0 \lsim 300$ GeV
and $M_{1/2} \lsim 190$ GeV. The bounds from charginos, neutralinos and
gluinos searches exclude all the values for the $M_0$ up to 800 GeV  where
 $M_{1/2}$ is less than 140 GeV while they exclude all the values of $M_0
\lsim 800$ GeV with $M_{1/2} \lsim 120$ GeV  when we assume 
non universal boundary conditions. These bounds are valid in all the
figures that follow in this article.

\begin{center}
\begin{tabular}{|c|c|c|c|c|}\hline
{\rm Particle} &  {\rm Bound} 
 & {\rm Assumptions} & {\rm Reference} \\[2mm] \hline \hline
$m_{\tilde{\chi}_1^0}$ &  31 (42)   &  all $M_0$ ($M_0\ge$ 500) and 
$\tan\beta \ge 2$ &
{\rm \cite{Abbiendi:1998rz}}\\ \hline
$m_{\tilde{\chi}_2^0}$ &  61 (72)   &  all $M_0$ ($M_0\ge$ 500)  and 
$\tan\beta \ge 2$ &
{\rm \cite{Abbiendi:1998rz}}\\ \hline
$m_{\tilde{\chi}_3^0}$ &  102   &    all $M_0$ &
{\rm \cite{Abbiendi:1998rz}}\\ \hline
$m_{\tilde{\chi}_4^0}$ &  127   &    & 
\cite{caso}\\ \hline \hline
$m_{\tilde{\chi}_1^\pm}$ & 84 (90.0)   &  
all $M_0\ge 100$  and $M_{1/2}\ge 100$ ($M_{1/2}\ge 150$) GeV
& {\rm \cite{Abbiendi:1998rz}}\\ \hline
$m_{\tilde{\chi}_2^\pm}$ &  99   &     & 
\cite{caso}\\ \hline \hline
$m_{\tilde{\nu}}$ &  43.1   &     & 
\cite{caso}\\ \hline \hline
$m_{\tilde{e}_R}$ &  84   &  for $m_{\tilde{\chi}_1^0}<50$
   & \cite{Abbiendi:1998ar} \\ \hline
$m_{\tilde{\mu}_R}$ &  80   &    & \cite{Abreu:1998jy} \\ \hline 
$m_{\tilde{\tau}_R}$ &  80   &     & \cite{Abreu:1998jy} \\ \hline \hline
$m_{\tilde{q}}$ & 250 &   & \cite{Abbott:1999xc} \\ \hline
$m_{\tilde{t}_1}$ & 83 (120) &   $\theta_{\tilde{t}}=56^o$ 
($\theta_{\tilde{t}}=0^0$) and $m_{\tilde{\chi}_1^0} < 50$ &
\cite{navas,de} \\ \hline
$m_{\tilde{b}_1}$ & 83 &   & \cite{Abbott:1999wt} \\ \hline
$m_{\tilde{g}}$ & 300 &   & \cite{Abbott:1999xc} \\ \hline \hline
$m_h$ & 78.8 & & \cite{Hocker:1999pf} \\ \hline
$m_A$ & 79.1 & & \cite{Hocker:1999pf} \\ \hline
$m_{H^\pm}$ & 60 & $0.97 < \tan\beta < 40.9$ & \cite{Abbott:1999mt} \\ \hline
\end{tabular}
\end{center}
{\footnotesize{Table I : Current experimental bounds on the masses of the SUSY 
and Higgs particles. The assumptions used and the sources are
also displayed.}}

\hspace*{0.5cm}

The direct bounds  from SUSY particles searches are depicted in
Fig.\ref{fig2} in the case of relatively large value of $\tan\beta=30$.
In the case of universal boundary conditions the upper
left area is forbidden by the requirement of the Charge and Color
Breaking minima (CCB).
Note that the full one loop corrections
to the effective potential have been included.  
In this area some of the (squared) squark or slepton  masses become negative
in the vicinity of  the electroweak scale, bound which is related to the
charged and color breaking minima one. In this case, either this pattern
of masses is ruled out or there must be new physics beyond 
the MSSM at or below that scale~\cite{Falk:1996cq}. Such a bound
does not exist if one breaks the universality by 
the pattern of   eq.(\ref{nonun}). However,  the 
bounds from gaugino's searches are stronger in the latter case
and in addition new bounds from the requirement that the LSP is
the lightest neutralino, arise.

In Figs.\ref{fig:squarks},\ref{fig:higgs} we present the 
resulting spectrum for the light squarks of the third generation,
the light charged slepton, and the light Higgs boson. The light stop mass
is lighter in the case of the non-universality and the 
opposite happens to be with the light bottom squark. This fact
is easily understood from eq.(\ref{nonun}). The light tau slepton 
mass turns out to be smaller in the case of where non-universal
boundary conditions are assumed. This is a renormalization group effect
and  we will discuss it below.
 The light sbottom squark
(squared) mass
is a function of the combination $m_{\tilde{Q}}^2+m^2_{\tilde{d}_R}$
which is larger than the combination  $m_{\tilde{Q}}^2+m^2_{\tilde{u}_R}$
which, ignoring the electroweak breaking effects, is the (squared) mass of the
light top squark.  However, this is only one part of the effect
of the non universal boundary conditions. There is another  one
which comes from the Renormalization Group analysis and affects
all the squarks and sleptons. Thus  every RGE for the (squared) soft 
SUSY breaking masses
contains a term\footnote{This term appears in the RGE of the 
soft supersymmetry breaking masses when the gauge group contains a $U(1)$
\cite{Martin:1994zk,yamada,Jack1,Jack2}.}
\begin{eqnarray}
16\pi^2\frac{d m_{\tilde{q}}^2}{d t} \ \subset \ {\cal A}_{\tilde{q}}~ g_1^2
\biggl \{ m_{H_2}^2-m_{H_1}^2 + {\rm Tr} \left ({\bf m}_{\tilde{Q}}^2  
-2 {\bf m}_{\tilde{U}^c}^2 +{\bf m}_{\tilde{D}^c}^2 -
{\bf m}_{\tilde{L}}^2 + {\bf m}_{\tilde{E}^c}^2 \right ) \biggr \} \;,
\end{eqnarray}
where ${\cal A}_{\tilde{q}}$ is a numerical factor, {\it i.e.,} in
the case of the selectron mass is ${\cal A}_{\tilde{E^c}}=\frac{6}{5}$.
This term is a multiplicative renormalised term in the absence of
threshold corrections but in this analysis where all the thresholds
in the RGE for the squark masses have been taken into account by
the mean of the ``theta'' function approximation \cite{Dedes:1996sb},
this is not the case. However, for universal boundary conditions this
 term vanishes at the GUT scale and its effect in the running of the
squark masses is rather small. If one assumes non-universal boundary 
conditions this term gives a major contribution to the RGE. Thus in our
case of eq,(\ref{nonun}) we obtain at the GUT scale,
\begin{eqnarray}
16\pi^2\frac{d m_{\tilde{q}}^2}{d t} \ \subset \
 {\cal A}_{\tilde{q}}~ g_1^2
~ 4 M_0^2
\end{eqnarray}
which affects dramatically the squark and slepton masses as
we can see from Fig.\ref{fig:squarks},\ref{fig:higgs}. Note, that in the case 
of the light Higgs boson mass the factors ${\cal A}$ come with
opposite sign, ${\cal A}_{H_1}=-{\cal A}_{H_2}=1/2$, in the running of 
the (squared) masses  $m_1^2=m_{H_1}^2+\mu^2$ and $m_2^2=m_{H_2}^2+\mu^2$
and thus we do not observe  significant effect {\it see,} 
Fig.\ref{fig:higgs}. However, it is worth to note that the 
light Higgs boson mass turns out to be larger by about 2-4 GeV 
in the case of the boundary conditions of eq.(\ref{nonun}).

In Fig.\ref{fig:lsp} we plot the masses of the lightest neutralinos
and charginos. We display only the results of the universal case
since there is only a small difference in the non-universal one.
However, some few GeV differences are enough to change the allowed 
or the excluded regions
displayed in Fig.(\ref{fig1},\ref{fig2}).  

In Fig.\ref{fig:as} we plot the resulting values of
the strong coupling $\alpha_s(M_Z)$, as function of the universal gaugino 
and squark masses $M_{1/2}$ and $M_0$. Threshold corrections affect
its value and thus we expect differences in the extracted values
in both cases. Indeed, we observe that in the case of 
the non universal boundary condition the $\alpha_s(M_Z)$ turns 
out to be 3\% 
smaller and thus closer  to its experimental value, $\alpha_s(M_Z)
=0.119 \pm 0.002$ \cite{caso}. In fact, for $M_{1/2}=450$ GeV and $M_0=800$ GeV
the theoretical result is 4$\sigma$ in the case of universal
and 2.5$\sigma$ in the case of non universal boundary conditions, away 
from the experimental result. For a rather light spectrum, 
$M_{1/2}=200$ GeV and $M_0=300$ GeV the theoretical observation 
with the boundary conditions of eq.(\ref{un}) is 6.5$\sigma$ far
from the experimental value while in the case of the boundary 
conditions of eq.(\ref{nonun}) is 5$\sigma$. In the case of 
large value of the $\tan\beta$ the value of $\alpha_s$ turns out
to be 0.001 larger than the case of low  values of $\tan\beta$
in both cases. Thus we conclude that in the case with non-universal boundary
conditions the extracted value of the strong coupling tends to
agree with the experimental data. At this point we should
say that the string boundary
conditions may affect the low energy results
in interesting ways. The uncertainties may be 
quite large because there may be additional
matter in the desert. The string threshold corrections have  been
discussed in detail in Ref.\cite{dienes}.

In Fig.\ref{fig:sin} the resulting effective leptonic weak mixing
angle is depicted as a function of $M_{1/2}$ with different values
of $M_0$ and $\tan\beta$. For the experimental value, 
$\sin^2_{eff}(lept)=0.23168\pm 0.00036$,
 we show the LEP one where only the $e$ and $\mu$ asymmetries have 
been taken into account \cite{Altarelli:1998xf}. The theoretical predictions 
are in agreement with the experimental data for moderate and large values
of $\tan\beta$. In the decoupling limit (where all the sparticles are heavy)
we obtain a single value of the $\sin_{eff}^2$=0.23135 (0.23150) for
$\tan\beta=$2 (30). These values are independent of the input values
at the GUT scale for the $M_0$, $M_{1/2}$ thanks to the decoupling 
of the SUSY particles (for more details see Ref.\cite{Dedes:1998hg}).
No significant differences have been obtained between the universal 
and non universal cases. 

The prediction of the W-boson pole mass is coming next.
We plot it in the Fig.\ref{fig:mw} for fixed values of $m_t=175$ GeV
and $A_0=0$ GeV.
We observe agreement with the experimental data both from
CDF ($80.405\pm 0.089$) and LEP ($80.427\pm 0.075$)~\cite{caso}.
 In the decoupling region, the W-boson pole
mass takes on values 80.397 (80.412) for $\tan\beta=30$ ($\tan\beta=2$),
and for  all the input values\footnote{Variation of the trilinear
coupling affects very slightly the results, {\it i.e.,} see for instance
Ref.\cite{Dedes:1998hg}}. We obtain large changes of the extracted
W-pole mass only in the case of small values of $M_0$ and $M_{1/2}$.

\section{Conclusions}

\begin{figure}
\centerline{\psfig{figure=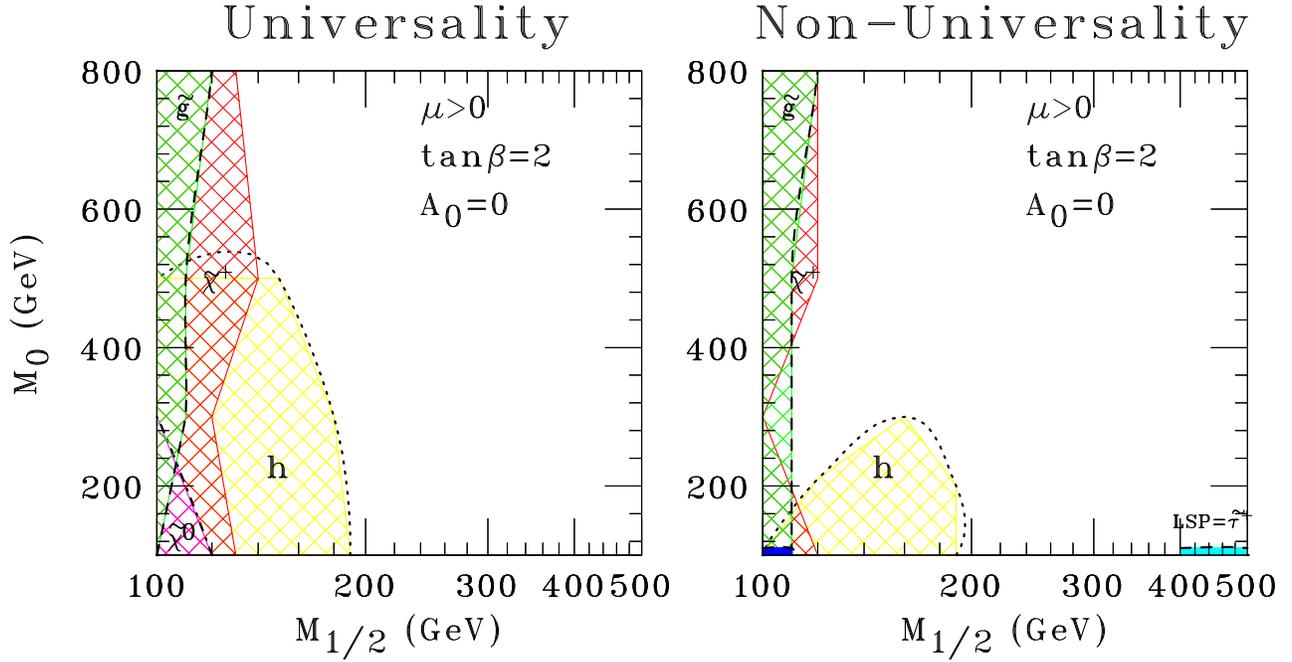,width=3.5in,angle=90}}
\caption{Excluded region in the $M_0$-$M_{1/2}$ plane from the experimental
bounds of the SUSY particles. We have chosen the values displayed in the
figure for the other
input parameters. The excluded regions are denoted with the shaded ones
and the particle which fails to pass the bound of Table I. Small shaded
regions in the non-universality case indicate excluded regions from
neutralinos and scalar $\tilde{\tau}$'s (left) and $\tilde{\tau}$  as
an LSP (right). }
\label{fig1}
\end{figure}

\begin{figure}
\centerline{\psfig{figure=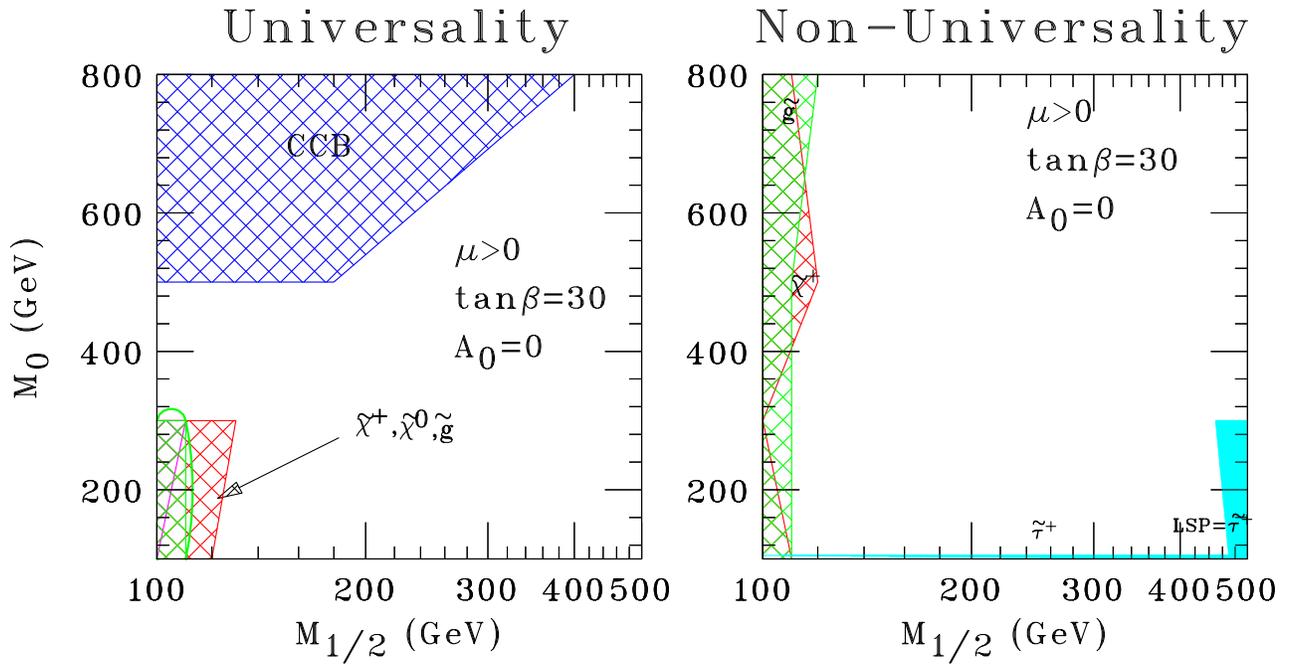,width=3.5in,angle=90}}
\caption{The same as Fig.\ref{fig1} with $\tan\beta=30$.}
\label{fig2}
\end{figure}

\begin{figure}
\centerline{\psfig{figure=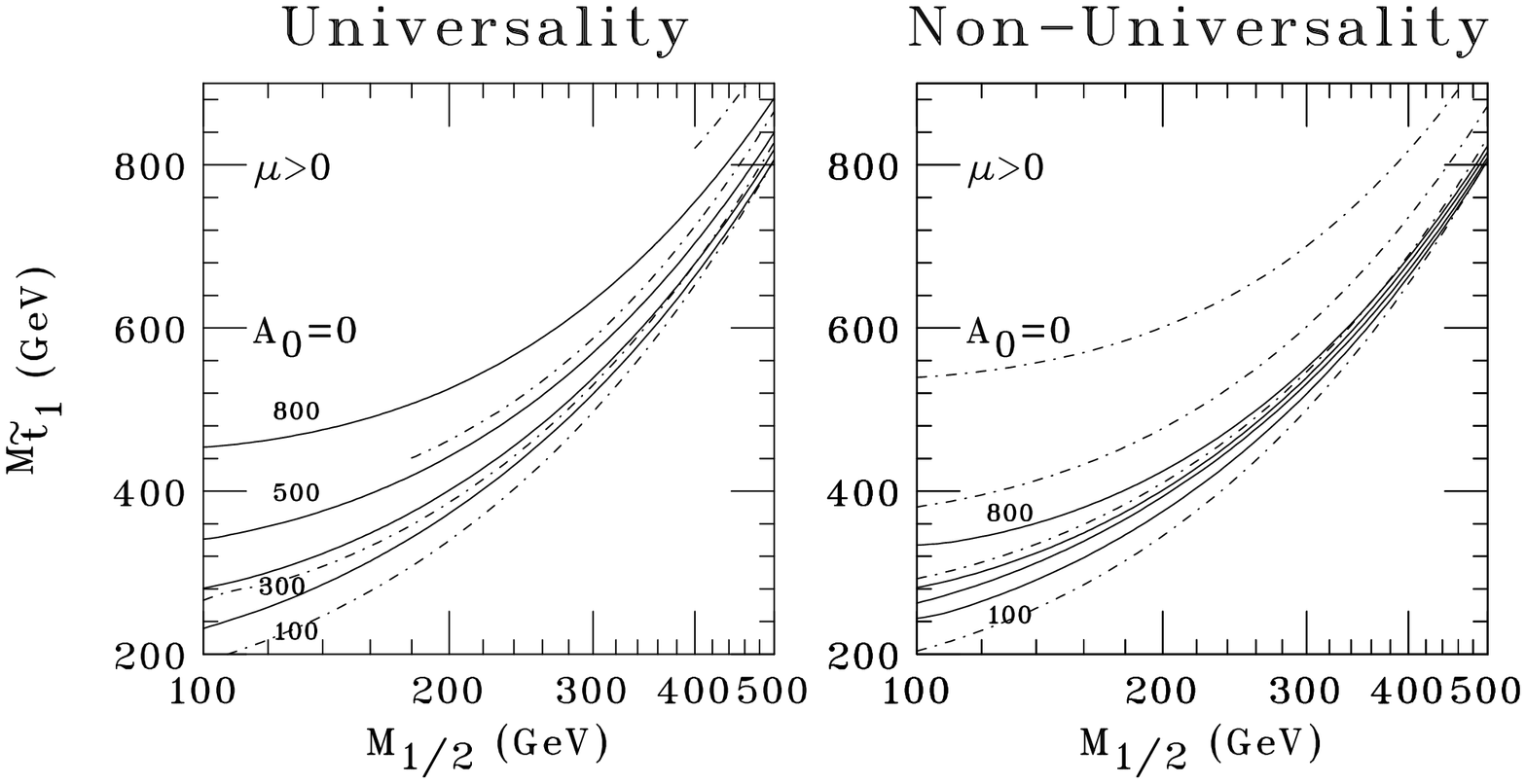,width=3.5in,angle=90}}
\centerline{\psfig{figure=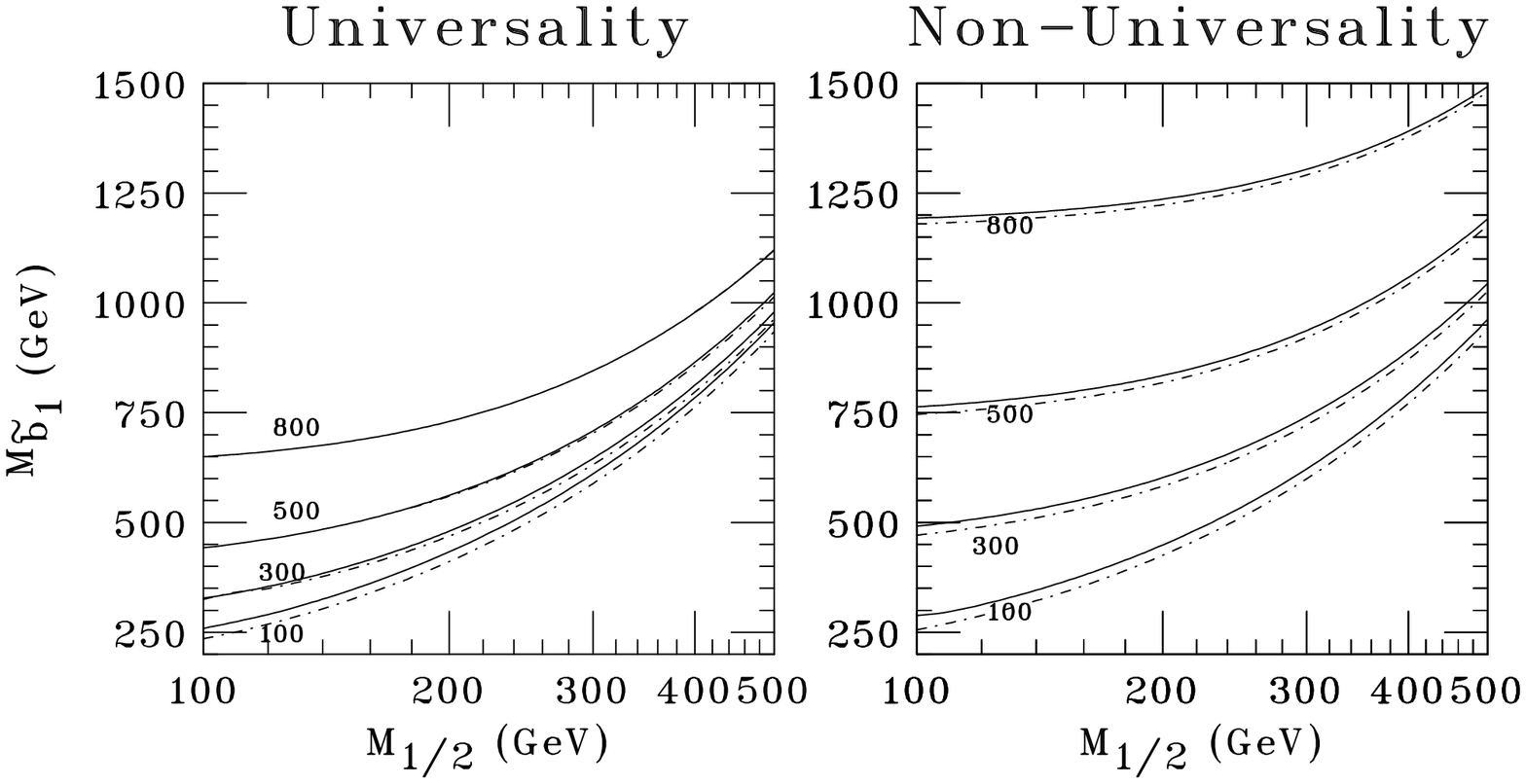,width=3.5in,angle=90}}
\caption{Predictions for the light top and bottom scalar quark
masses as a function of $M_{1/2}$ for different values of $M_0$ (displayed)
and $\tan\beta$=2(solid) 30(dot-dashed) values.}
\label{fig:squarks}
\end{figure}

\begin{figure}
\centerline{\psfig{figure=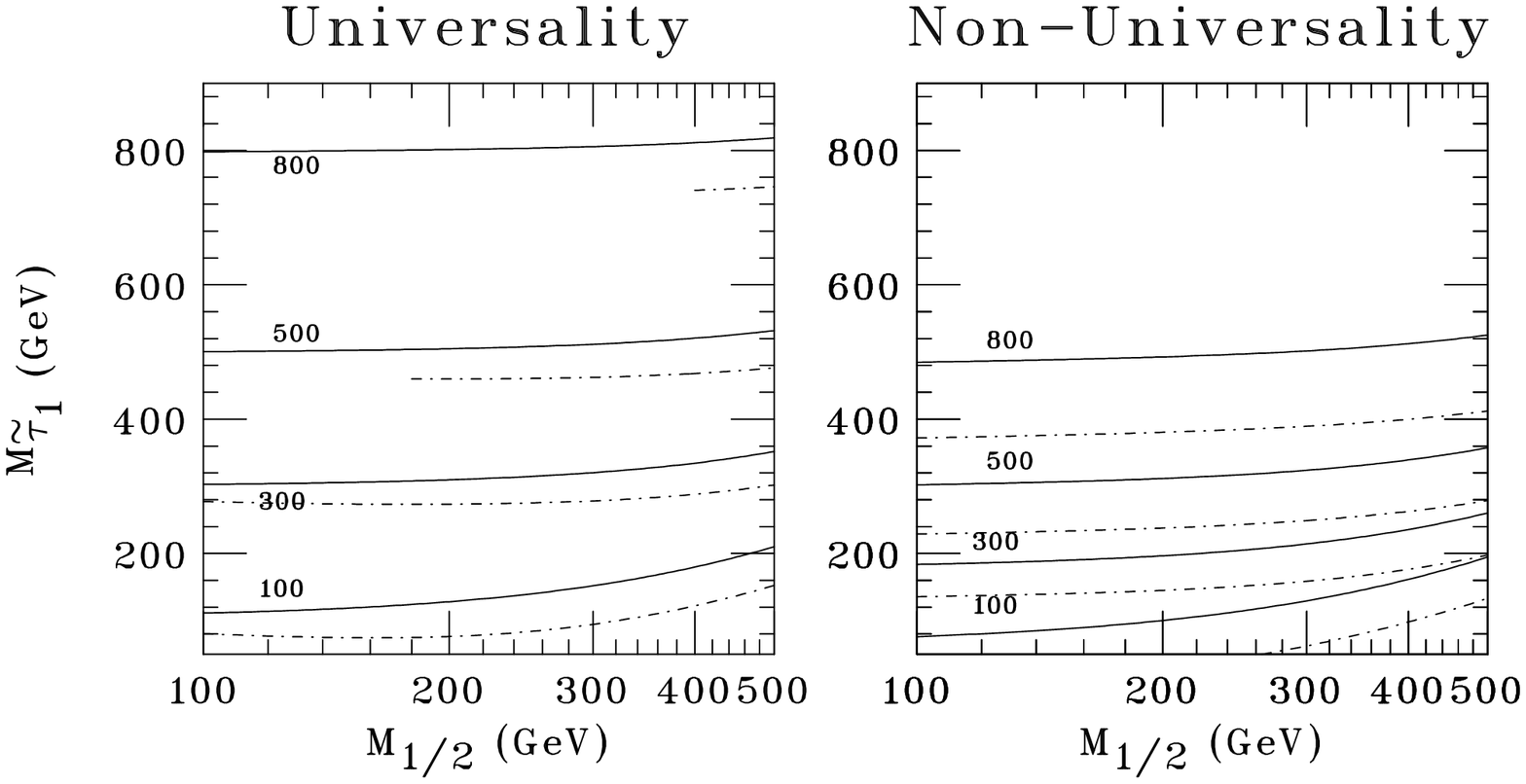,width=3.5in,angle=90}}
\centerline{\psfig{figure=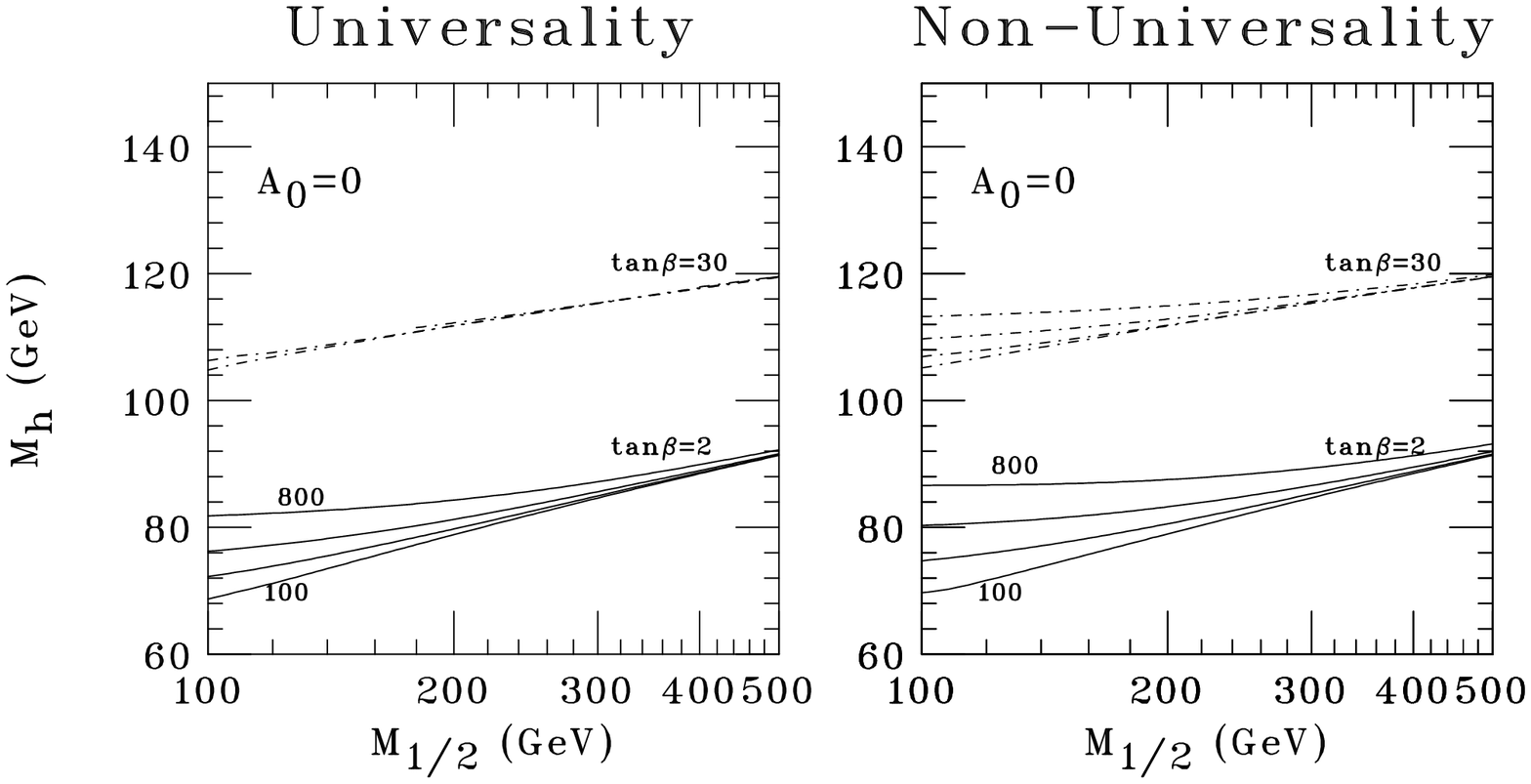,width=3.5in,angle=90}}
\caption{Predictions for the light tau scalar lepton and for 
the light Higgs boson 
masses as a function of $M_{1/2}$ for different values of $M_0$ (displayed)
and $\tan\beta$=2(solid) 30(dot-dashed) values.}
\label{fig:higgs}
\end{figure}

\begin{figure}
\hbox{\psfig{figure=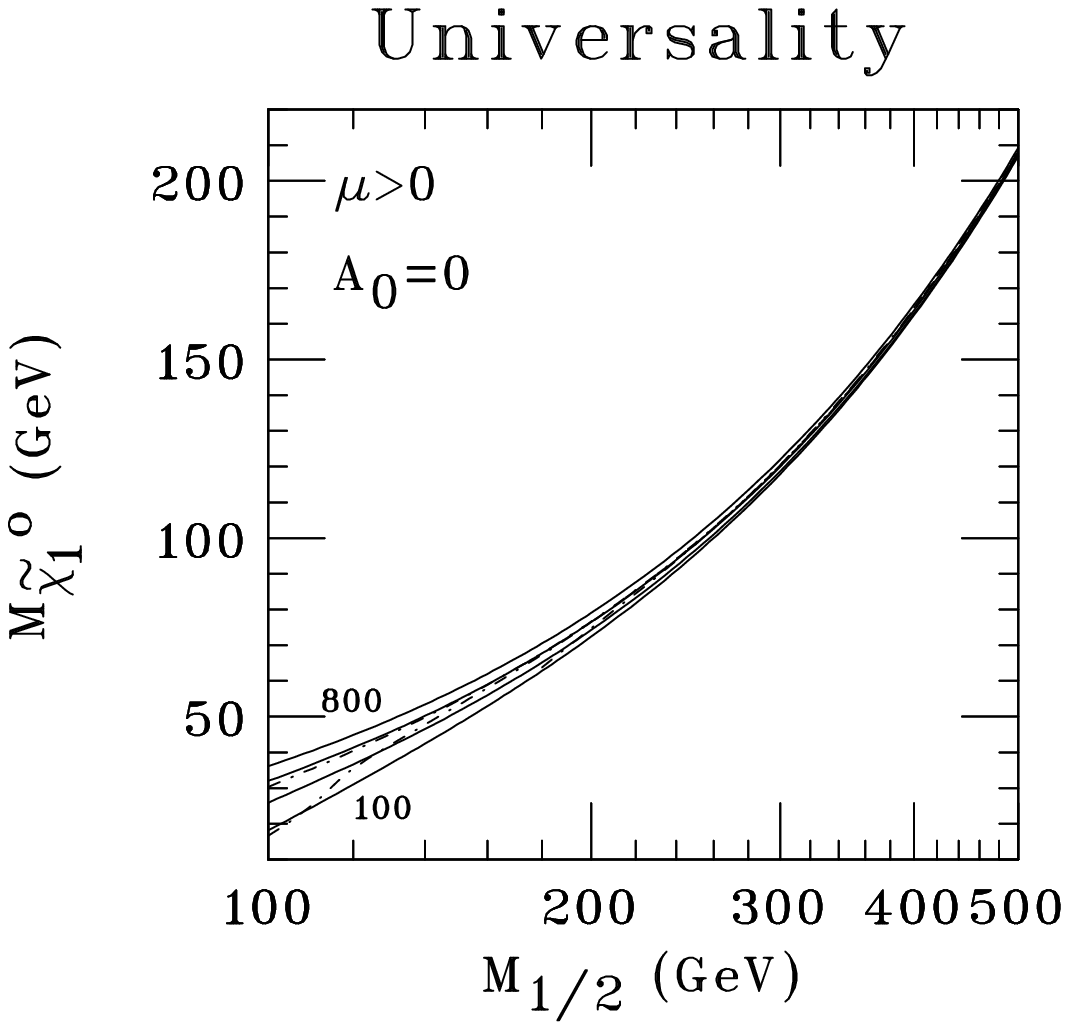,width=3.0in,angle=90}
\psfig{figure=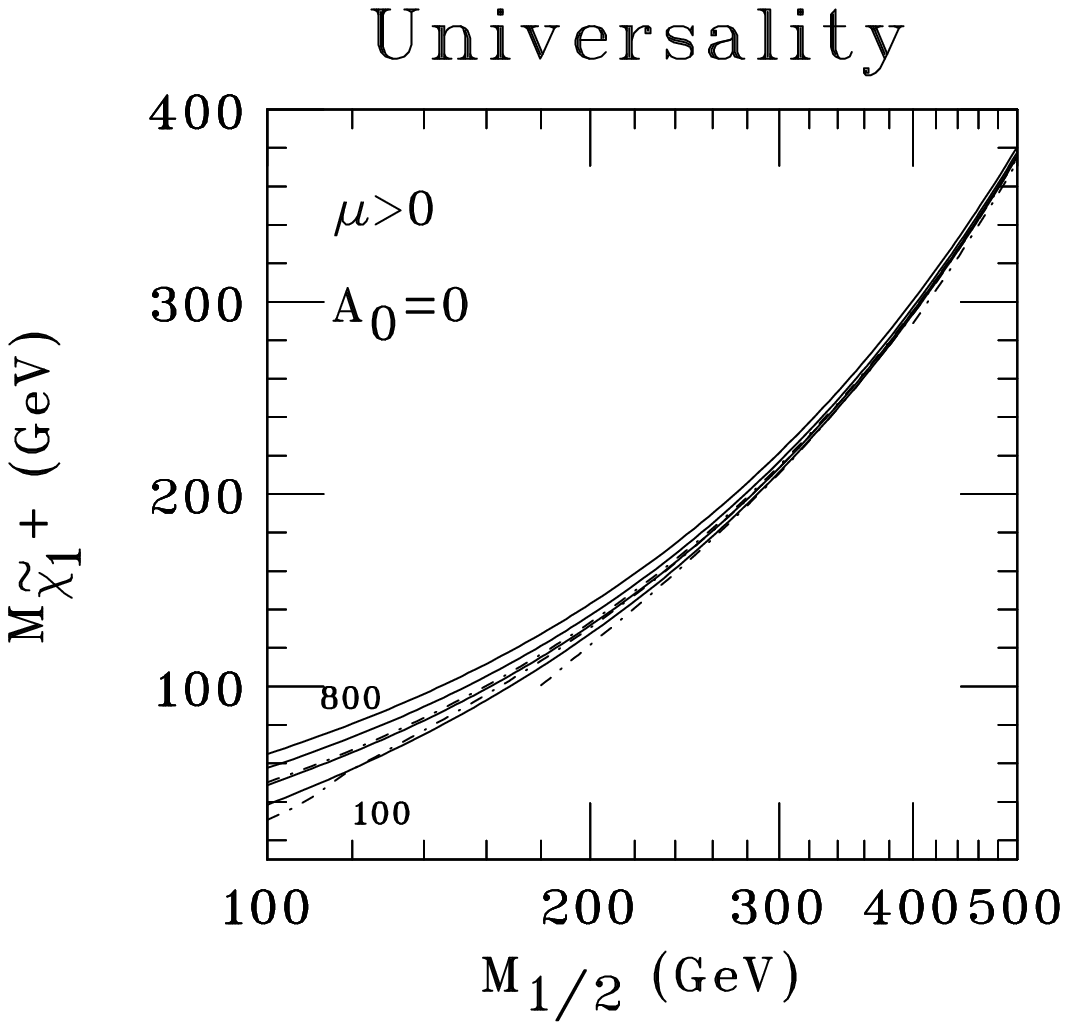,width=3.0in,angle=90}}
\caption{Predictions for the lightest neutralino (LSP) and chargino   
masses as a function of $M_{1/2}$ for different values of $M_0$ (displayed)
and $\tan\beta$=2(solid) 30(dot-dashed) values.}
\label{fig:lsp}
\end{figure}

\begin{figure}
\centerline{\psfig{figure=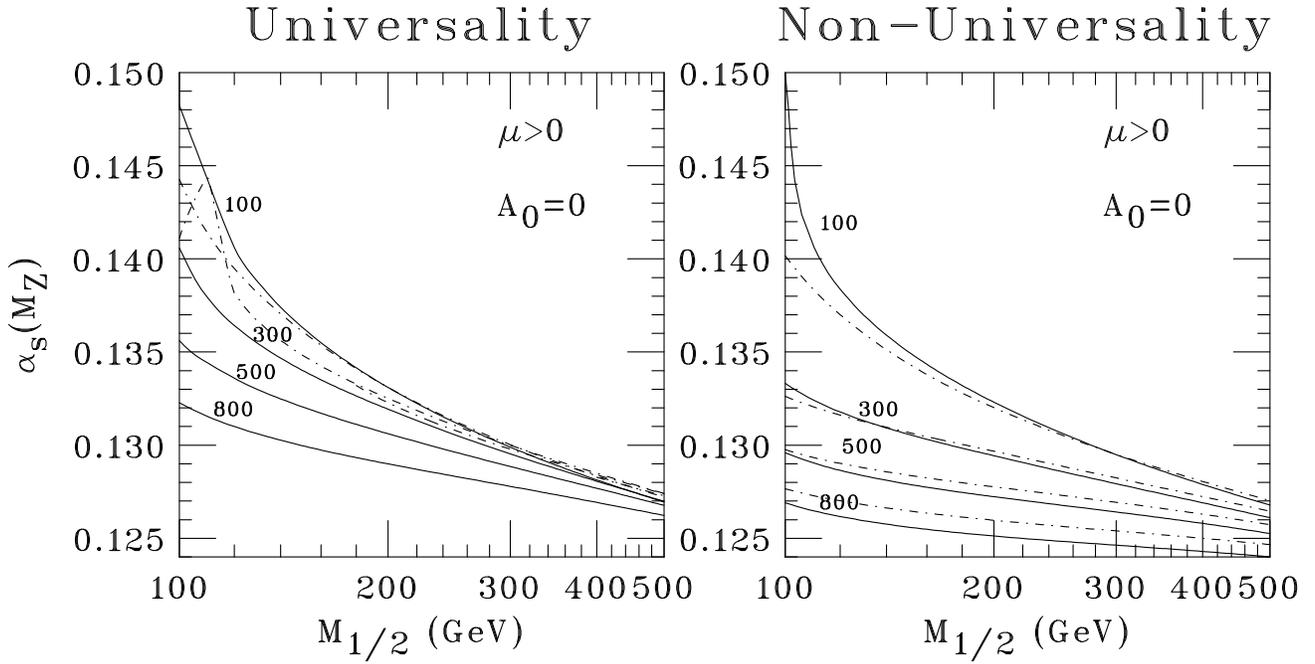,width=3.5in,angle=90}}
\caption{Resulting values of the strong coupling $\alpha_s(M_Z)$ as
a function of $M_0$ for different values of the $M_0$. The top
quark mass 175 GeV is assumed. We display results for both regions of  small 
$\tan\beta=2$ (solid lines) and large $\tan\beta=30$ (dot-dashed) lines). }
\label{fig:as}
\end{figure}

\begin{figure}
\centerline{\psfig{figure=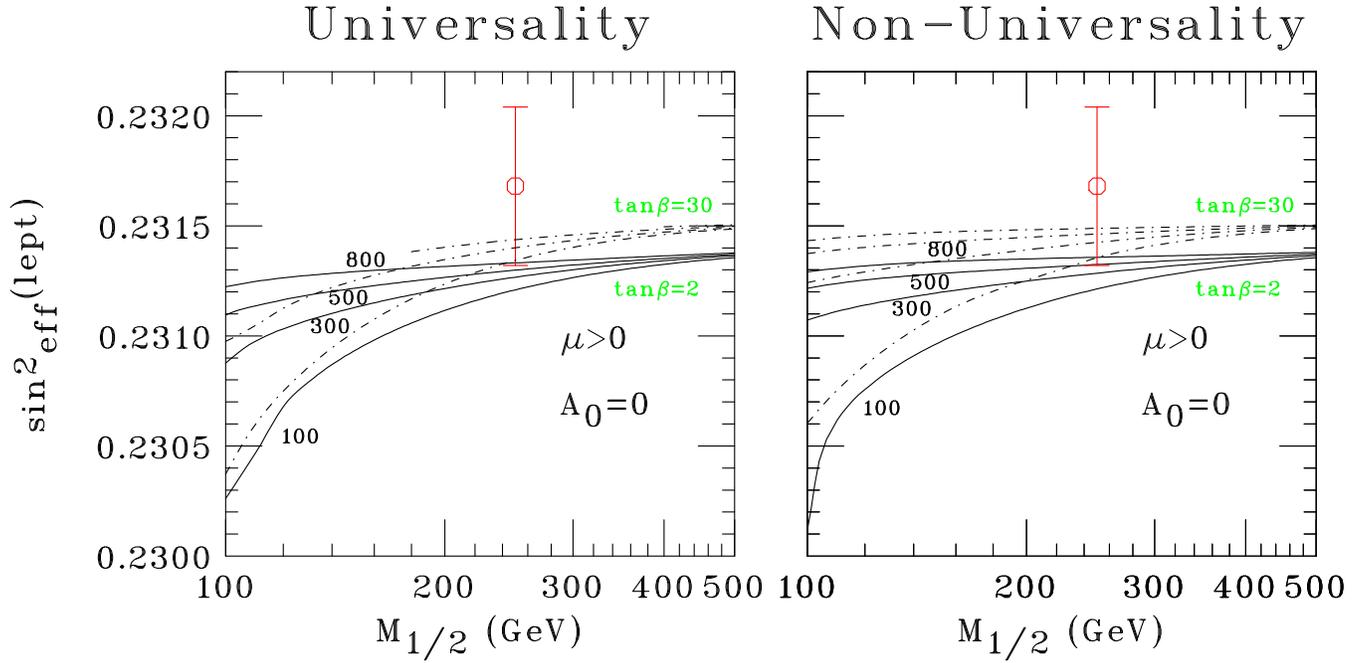,width=3.5in,angle=90}}
\caption{Resulting values of the effective (leptonic)
weak mixing angle, $\sin^{2}_{eff}$(lept) as
a function of $M_{1/2}$
 for different values of the $M_0=$100,300,500,800 GeV and
two different values of $\tan\beta=$2,30 (dot-dashed lines). 
The LEP experimental value is also 
shown.}
\label{fig:sin}
\end{figure}

\begin{figure}
\centerline{\psfig{figure=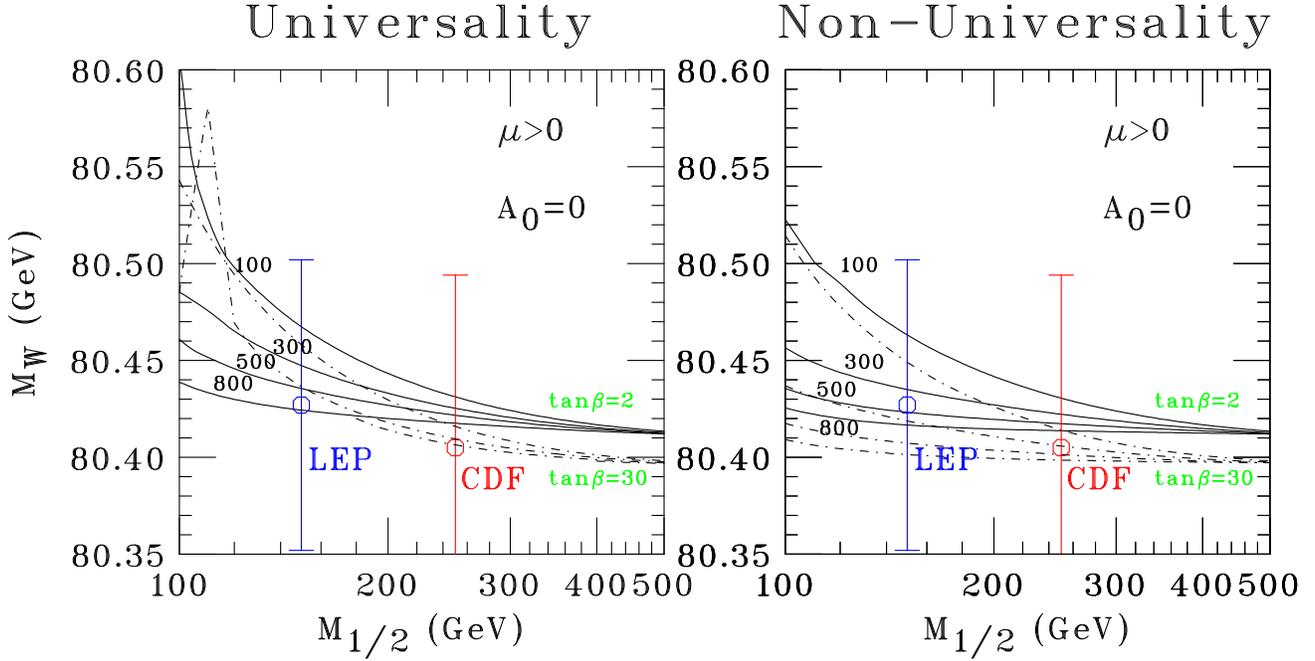,width=3.5in,angle=90}}
\caption{Resulting values of the physical W-pole mass, $M_W$
 as
a function of $M_{1/2}$
 for different values of the $M_0=$100,300,500,800 GeV and
two different values of $\tan\beta=$2,30 (dashed lines). The 
experimental CDF and D\O ~values are also displayed.}
\label{fig:mw}
\end{figure}

In this article we aimed at achieving two goals :
 {\bf 1.} To discuss how the anomalous U(1)
charges can be useful to study different
string compactifications,  in concrete 
superstring models. The idea here is 
to show that irrespective of what we don't 
know about the mechanism of supersymmetry breaking
the signature of the anomalous U(1) charges
will still provide useful information. 
{\bf 2}. To analyse the implications of
the boundary conditions eqs.(\ref{un},\ref{nonun}) which are
given in  Ref~\cite{fp2}. We find that non-universality relaxes 
some of the experimental bounds. For example, in the case 
of $\tan\beta=2$ the mass of the light Higgs is increasing 
by 2-4 GeV ({\it see} Figs.(\ref{fig1},\ref{fig:higgs}) ). For 
large values of $\tan\beta=30$ the constraint from dangerous
Charge and Color breaking minima directions is removed in
the case of the non-universal boundary conditions of  
eq.(\ref{nonun}) but a new constraint (this is when the  LSP becomes
a charged slepton) appears in the 
region of large $M_{1/2}$ and small $M_0$ ({\it see} Fig.(\ref{fig2}) ).
 When non-universal boundary
conditions are assumed the region $M_{1/2}\lsim 110$ GeV is excluded
for every value of $M_0\lsim 800$ GeV and every value of the $\tan\beta$
between 2 and 30. Large differences in the mass of the lightest 
top and  bottom squarks as well of tau slepton  about 10\% 
to 100\% 
are obtained ({\it see} Figs.(\ref{fig:squarks}) ). Lightest charginos and
neutralinos remain unchanged and here we display the predictions
for their masses only in the case of universal boundary conditions
({\it see} Figs.(\ref{fig:lsp}) ).  One can derive immediately the 
bounds on the boundary conditions, $M_0$ and $M_{1/2}$ either by comparing
the graphs with the Table I or by looking the Figs.(\ref{fig1},\ref{fig2}).
We derive also the predictions on the strong QCD coupling,
effective weak mixing angle and W-pole mass, of the two models
by taking into account all the SM and SUSY threshold corrections.
We find that the extracted value of $\alpha_s(M_Z)$ turns out
to be 3\% 
smaller in the case of the non-universal boundary conditions 
({\it see} Figs.(\ref{fig:as}) ) .
In addition, with squark masses up to 1 TeV the value
of $\alpha_s$ is $2.5\sigma$ far from its  experimental value. 
This discrepancy can be removed from string/GUT threshold corrections
which have not been taken into account here.  No significant 
changes are observed for the predicted values of $\sin^2_{eff}(lept)$
between the two models ({\it see} Figs.(\ref{fig:sin}) ) . 
The predicted pole mass of the W-gauge boson
is in agreement with the data in both models although it prefers
non-universal boundary conditions in the region of light $M_{1/2}$ and $M_0$
({\it see} Figs.(\ref{fig:mw}) ).

\vspace*{1cm}

{\bf Acknowledgements}

AEF thanks the CERN theory division for hospitality while
part of this work was conducted.
This work was supported in part by 
DOE Grant No.\ DE-FG-0294ER40823. A.D is supported from Marie 
Curie Research Training Grants
ERB-FMBI-CT98-3438.

\bibliographystyle{unsrt}

\begin{thebibliography}{99}

\bibitem{revamp} I. Antoniadis, J. Ellis, J. Hagelin and D.V. Nanopoulos,
                        \PLB{231}{89}{65};
                 J.L. Lopez, D.V. Nanopoulos and K. Yuan,
                        \NPB{399}{93}{654}.
\bibitem{fny} A.E. Faraggi, D.V. Nanopoulos, and K. Yuan, \NPB{335}{90}{347}.
\bibitem{alr} I. Antoniadis, G.K. Leontaris and J. Rizos, \PLB{245}{90}{161};\\
                G.K. Leontaris, \PLB{372}{96}{212};\\
                 G.K. Leontaris and J. Rizos, hep-th/9901098.
\bibitem{eu}     A.E. Faraggi, \PLB{278}{92}{131}; \NPB{387}{92}{239}.
\bibitem{top}    \AEF, \PLB{274}{92}{47}; \PLB{339}{94}{223}.
\bibitem{cfn} G.B. Cleaver, A.E. Faraggi and D.V. Nanopoulos, 
  \PLB{455}{99}{135}; hep-ph/9904301.
\bibitem{ccf}  X.G. Wen and E. Witten, \NPB{265}{85}{651};\\
               G.G. Athanasiu, J.J. Atick, M. Dine, and W. Fischler,
               \PLB{214}{88}{55};\\
               A.N. Schellekens, \PLB{237}{90}{363};\\
               J. Ellis, J.L. Lopez and D.V. Nanopoulos, \PLB{247}{90}{257};\\
               S. Chang, C. Coriano and A.E. Faraggi, \PLB{397}{97}{76};
               \NPB{477}{96}{65}.
\bibitem{lykken} S. Chaudhoury, G. Hockney and J. Lykken, 
                                                \NPB{469}{96}{357};\\
                 G.B. Cleaver {\it et. al.} , \NPB{525}{98}{3}; 
                                                \NPB{545}{98}{47};
                                                \PRD{59}{99}{055005};
                                                \PRD{59}{99}{115003}.
\bibitem{ben} K. Benakli, J. Ellis and D.V. Nanopoulos, \PRD{59}{99}{047301};\\
              A.E. Faraggi, K.A. Olive and M. Pospelov, hep-ph/9906345.
\bibitem{gauginocond} M. Dine, R. Rohm, N. Seiberg and E. Witten,
                                        \PLB{156}{85}{55};\\
        J.P. Derendinger, L.E. Ibanez and H.P. Nilles, \PLB{155}{85}{65};\\
        H.P. Nilles, \PLB{115}{82}{193}.
\bibitem{DDSB} V. Kaplunovsky and J. Louis, \PLB{306}{93}{269}.
\bibitem{GMSB}
        M. Dine, A.E. Nelson, Y. Nir and Y. Shirman, \PRD{53}{96}{2658}.
\bibitem{fh1} P. Fayet, \NPB{90}{75}{104};\\
               I. Antoniadis, John Ellis, A.B. Lahanas and D.V. Nanopoulos,
               \PLB{241}{90}{24};\\
              A.E. Faraggi, E. Halyo, \IJMP{11}{96}{2357};\\
                G. Dvali and A. Pomarol, \PRL{77}{96}{3728};\\
                P. Binetruy and E. Dudas, \PLB{389}{96}{503}.
                R. Mohapatra and A. Riotto, \PRD{55}{97}{1138};
               N. Arkani--Hamed, M. Dine and S.P. Martin, \PLB{431}{98}{239};\\
T. Barreiro, B. de Carlos, J.A. Casas and J.M. Moreno, \PLB{445}{98}{82};\\
N. Irges, \PRD{59}{99}{115008}.

\bibitem{fp2} A.E. Faraggi and J.C. Pati, \NPB{526}{98}{21}.

\bibitem{il} L. Ibanez and D. Lust, \NPB{382}{92}{305}.
\bibitem{cf1} G.B. Cleaver and A.E. Faraggi, \IJMP{14}{99}{2335};\\
  A.E. Faraggi, \PLB{426}{98}{315}; hep-ph/9807341.
\bibitem{fff}  H. Kawai, D.C. Lewellen, and S.-H.H. Tye, \NPB{288}{87}{1};\\
               I. Antoniadis, C. Bachas, and C. Kounnas,
                                \NPB{289}{87}{87};\\
               I. Antoniadis and C. Bachas, \NPB{298}{88}{586}.
\bibitem{cvetic} L. Dixon, E. Martinec, D. Friedan and S. Shenker,
                        \NPB{282}{87}{13};\\
                 M. Cvetic, \PRL{59}{87}{2829}.
\bibitem{kln}    S. Kalara, J.L. Lopez and D.V. Nanopoulos, \NPB{353}{91}{650}.
\bibitem{nahe}   A.E. Faraggi and D.V. Nanopoulos, \PRD{48}{93}{3288};\\
                 \AEF, \NPB{407}{93}{57}; hep-th/9511093; hep-th/9708112.
\bibitem{foc}    A.E. Faraggi, \PLB{326}{94}{62}.
\bibitem{price} I. Antoniadis, J. Ellis, S. Kelley and D.V. Nanopoulos,
                        \PLB{272}{91}{31}. 


\bibitem{matter}
      M.K. Gaillard and R. Xiu, \PLB{296}{92}{71};\\  
      A.E. Faraggi, \PLB{302}{93}{302};\\
      S.P. Martin and P. Ramond, \PRD{51}{95}{6515}.

\bibitem{witten} E. Witten, \NPB{471}{96}{135}.

\bibitem{kelley} A.E. Faraggi, S. Kelley, J. Hagelin and D.V. Nanopoulos, 
  \PRD{45}{92}{3272};\\
S.P. Martin and P. Ramond, \PRD{48}{93}{5365};\\
Y. Kawamura, H. Murayama and M. Yamaguchi, \PLB{324}{94}{52};\\
J.L. Feng, M.E. Peskin, H. Murayama and X. Tata, \PRD{52}{95}{1418};\\
Y. Kawamura, T. Kobayashi and T. Komatsu, \PLB{400}{97}{284}.

\bibitem{Dedes:1998hg}
A.~Dedes, A.B.~Lahanas and K.~Tamvakis, \PRD{59}{99}{015019}.

\bibitem{Dedes:1997wc}
A.~Dedes, A.B.~Lahanas, J.~Rizos and K.~Tamvakis, \PRD{55}{97}{2955}.

\bibitem{Faraggi:1994qb}
A.E.~Faraggi and B.~Grinstein, \NPB{422}{94}{3}.

\bibitem{Bagger:1995bw}
J.~Bagger, K.~Matchev and D.~Pierce, \PLB{348}{95}{443}.

\bibitem{Lahanas:1995dj}
A.B.~Lahanas and K.~Tamvakis, \PLB{348}{95}{451}.

\bibitem{Dedes:1996sb}
A.~Dedes, A.B.~Lahanas and K.~Tamvakis, \PRD{53}{96}{3793}.

\bibitem{Abbiendi:1998rz}
G.~Abbiendi {\it et al.}
[OPAL Collaboration],
Eur. Phys. J. {\bf C8} (1999) 255.

\bibitem{caso} 
C.~Caso {\it et al.,} The European Physical Journal {\bf C3} (1998) 1.

\bibitem{Abbiendi:1998ar}
G.~Abbiendi {\it et al.}
[OPAL Collaboration], hep-ex/9808036.

\bibitem{Abreu:1998jy}
P.~Abreu {\it et al.}
[DELPHI Collaboration], \PLB{444}{98}{491}.

\bibitem{Abbott:1999xc}
B.~Abbott {\it et al.}
[D0 Collaboration], hep-ex/9902013.

\bibitem{navas}
S. Navas-Concha, 'MSSM searches at LEP', talk given at
`SUSY~98', Keble College, Oxford, UK, July 11-17 1998.

\bibitem{de}
J. Valls, `MSSM and Higgs Search at the Tevatron', XXIX ICHEP'98,
Vancouver Conference, July 1998;\\
K. De, 'MSSM searches at the Tevatron', talk given at
`SUSY~98', Keble College, Oxford, UK, July 11-17 1998.


\bibitem{Abbott:1999wt}
B.~Abbott {\it et al.}
[D0 Collaboration], hep-ex/9903041.

\bibitem{Hocker:1999pf}
A.~Hocker, hep-ex/9903024.


\bibitem{Abbott:1999mt}
B.~Abbott {\it et al.}
[D0 Collaboration], hep-ex/9902028.


\bibitem{Falk:1996cq}
T.~Falk, K.A.~Olive, L.~Roszkowski and M.~Srednicki,
\PLB{367}{96}{183}.

\bibitem{Martin:1994zk}
S.P.~Martin and M.T.~Vaughn, \PRD{50}{94}{2282}.

\bibitem{yamada}
Y.~Yamada, \PRD{50}{94}{3537}.

\bibitem{Jack1}
I.~Jack and D.R.~Jones, \PLB{349}{95}{294}.

\bibitem{Jack2}
I.~Jack, D.R.~Jones and K.L.~Roberts, \NPB{455}{95}{83}.

\bibitem{dienes}
K.R.~Dienes and A.E.~Faraggi, \NPB{457}{95}{409}.

\bibitem{Altarelli:1998xf}
G.~Altarelli, hep-ph/9811456.

\end{thebibliography}

\newpage

\vfill
\eject

\end{document}